\documentclass[sigconf,review=false,anonymous=false,authordraft=false]{acmart}
\AtBeginDocument{%
  \providecommand\BibTeX{{%
    \normalfont B\kern-0.5em{\scshape i\kern-0.25em b}\kern-0.8em\TeX}}}

\copyrightyear{2024}
\acmYear{2024}
\setcopyright{rightsretained}

\acmConference[SIGIR '24]{Proceedings of the 47th International ACM SIGIR Conference on Research and Development in Information Retrieval}{July 14--18, 2024}{Washington, DC, USA.}
\acmBooktitle{Proceedings of the 47th Int'l ACM SIGIR Conference on Research and Development in Information Retrieval (SIGIR '24), July 14--18, 2024, Washington, DC, USA}

\usepackage{comment}
\usepackage{algorithm}
\usepackage{algorithmic}
\usepackage{xcolor}
\usepackage{xspace}
\usepackage{multirow, diagbox}
\usepackage{pifont}
\usepackage{enumitem}
\usepackage{balance}
\usepackage{cleveref}
\usepackage{placeins}

\newcommand{\framework}{PAG\xspace}
\begin{document}

\title{GD-RIPOR: Globally-Guided Constrained Beam Search for Generative Information Retrieval}
\title{Planning Ahead in Generative Retrieval: Guiding Autoregressive Generation through Simultaneous Decoding}


\author{Hansi Zeng}
\affiliation{%
  \institution{University of Massachusetts Amherst}
  % \streetaddress{P.O. Box 1212}
  % \city{Amherst}
  % \state{MA}
  \country{United States}
  % \postcode{01003}
}
\email{hzeng@cs.umass.edu}

\author{Chen Luo}
\affiliation{%
  \institution{Amazon}
  % \streetaddress{P.O. Box 1212}
  % \city{Amherst}
  % \state{MA}
  \country{United States}
  % \postcode{01003}
}
\email{cheluo@amazon.com}

\author{Hamed Zamani}
\affiliation{%
  \institution{University of Massachusetts Amherst}
  % \streetaddress{P.O. Box 1212}
  % \city{Amherst}
  % \state{MA}
  \country{United States}
  % \postcode{01003}
}
\email{zamani@cs.umass.edu}

\renewcommand{\shortauthors}{Hansi Zeng et al.}

\begin{abstract}
This paper introduces PAG--a novel optimization and decoding approach that guides autoregressive generation of document identifiers in generative retrieval models through simultaneous decoding. To this aim, PAG constructs a set-based and sequential identifier for each document. Motivated by the bag-of-words assumption in information retrieval, the set-based identifier is built on lexical tokens. The sequential identifier, on the other hand, is obtained via quantizing relevance-based representations of documents.  Extensive experiments on MSMARCO and TREC Deep Learning Track data reveal that PAG outperforms the state-of-the-art generative retrieval model by a large margin (e.g., $15.6\%$ MRR improvements on MS MARCO), while achieving $22\times$ speed up in terms of query latency.

\end{abstract}






\newcommand{\xmark}{\text{\ding{55}}}

\maketitle

\section{Introduction}
Generative Retrieval (GR), also referred to as differentiable search index, provides a novel paradigm for information retrieval, diverging from the traditional ``index-then-retrieve'' approach employed in sparse and dense retrieval models \cite{DPR,BM25,snrm,colbert,splade}. In GR, each document is first assigned a unique document identifier (DocID); then, a generative retrieval model, often based on a large language model (LLM), is trained to generate relevant DocIDs in response to a query \cite{DSI,dsi++,RIPOR,NCI,DSI-QG}. A distinct property of GR models is their capacity to consolidate the corpus information within their parameters, which makes their integration into other generation tasks that benefit from information retrieval differentiable and seamless \cite{RIPOR}. Important examples of such applications include knowledge-intensive text generation \cite{REML,guu2020retrieval,Lewis2020RetrievalAugmentedGF,fid-light,Salemi:2023:KIVQA} and personalized generation \cite{salemi2024lamp,salemi2024optimization}.

Each DocID in generative retrieval often consists of a sequence of tokens. Hence, they generate DocIDs autoregressively; meaning that they generate one token at a time, conditioned on the query encoding and the previously generated tokens. Borrowed from the language modeling literature, the (constrained) beam search algorithm  \cite{DSI,RIPOR,NCI,dsi++,DSI-QG} is used for generation during inference. However, unlike language generation where multiple equally acceptable outputs exist, each relevant document in generative retrieval is represented with only one identifier. Therefore, since beam search is a local search algorithm that tends to get stuck in local optima \cite{Zhou2005BeamStackSI,Stahlberg2019OnNS}, if all prefixes of this identifier do not survive the pruning process of beam search, there is no way to recover and the GR model would fail at retrieving the corresponding relevant document. 
Even though RIPOR \cite{RIPOR}---the current state-of-the-art generative retrieval model---achieves substantial improvements by emphasizing on accurate generation of DocID prefixes during training, our experiments show that many relevant DocIDs still exist that cannot survive beam search pruning in RIPOR. According to results presented in Figure \ref{fig:RIPOR_perf_vs_bs}, we observe that increasing the beam size would significantly affect the retrieval effectiveness, and even using a large beam size, e.g., 1000, still cannot meet the brute force decoding performance where every document in the corpus is scored (see Section~\ref{sec:motivation} for more details). 

Motivated by these findings, we propose \framework--a novel optimization and decoding approach that guides autoregressive generation through an efficient simultaneous DocID decoding for approximating document-level scores. In other words, each DocID consists of a set-based and a sequential identifier. \framework first decodes the set-based identifier, in which token ordering does not matter thus can be done in a single decoding step for approximating document-level scores. \framework then continues decoding the sequential identifier conditioned on the previous generations. Our hypothesis is that conditioning autoregressive decoding on document-level scores produced by simultaneous (i.e., set-based) decoding reduces the likelihood of a relevant prefix to be pruned by (constrained) beam search. Therefore, we revisit both optimization and decoding of generative retrieval models according to this hypothesis.

Inspired by the effectiveness of bag-of-words assumption in many existing retrieval models \cite{BM25,doc2query,SprckJones2021ASI,tf-idf,Ponte1998ALM,Zhai2001ModelbasedFI}, we construct our set-based document identifiers based on lexical tokens. Following \citet{RIPOR}, we also use residual quantization (RQ) over the relevance-based representations produced for each query and document by our GR model to form the sequential identifiers. We suggest a three-stage optimization pipeline, one for set-based DocID generation, one for sequential DocID generation, and one end-to-end training for joint set-based and sequential generation.

We conduct our evaluation on standard large-scale passage retrieval benchmarks including MSMARCO \cite{MSMARCO} and TREC Deep Learning Track Data from 2019 and 2020 \cite{Trec-19,Trec-20}, in which the corpus consists of 8.8 million passages. Compared to the current state-of-the-art generative retrieval model, i.e., RIPOR \cite{RIPOR}, \framework demonstrates $15.6\%$ relative improvement in terms of MRR@10 on MSMARCO Dev set and $12.3$\% and $10.9\%$ improvements in terms of NDCG@10 on TREC-DL 2019 and 2020, respectively. This is while \framework uses a $10\times$ smaller beam size, resulting in $22 \times$ improvement in terms of query latency when using a single A100 GPU for inference. 
% ( the improvement also partially benefits from using $4\times$ shorter sequential DocID compared to the original RIPOR).\hansi{check the sentence for query latency improvement} \hamed{i don't this is necessary for intro, but you can explain more when discussing efficiency results in experiments} 
Extensive ablation studies and analysis demonstrate the impact of the decisions we made in designing the \framework framework.
Even though the goal is not to compare with non-generative retrieval models, our experiments demonstrate improvements over several effective dense retrieval models. For instance, compared to TAS-B \cite{TAS-B}, RocketQA \cite{rocketqa}, and TCT-ColBERT \cite{tct-colbert}, \framework achieves $11.9\%$, $4.1\%$, and $14.9\%$ MRR@10 improvements on the MSMARCO Dev set, respectively. Another significant advantage of \framework over dense retrieval models is its memory efficiency. For example, it requires $7.7\times$ less memory to index the entire corpus (8.8 million passages) compared to single-vector dense retrieval models.

To improve reproducibility of this work and foster research in generative retrieval, we open-source our codebase and release trained model parameters at: \url{https://github.com/HansiZeng/PAG}.

\vspace{-.35cm}
\section{Related Work}
\underline{\textbf{Classic Neural IR Models}}:
With the emergence of large language models (LLMs) \cite{BERT,roberta,T5,instruct-gpt,flan-t5} and large-scale information retrieval datasets \cite{NQ,MSMARCO}, neural-based IR models have demonstrated superior results over the traditional lexical-matching models, such as BM25 \cite{BM25}. In general, these IR models can fall into three categories: (1) cross-encoder models \cite{BERT4rerank,rankT5,Pradeep2021TheED}, (2) dense retrieval models \cite{MarginMse,TAS-B,tct-colbert,DPR,Khattab2020ColBERTEA,CL-DRD}, and (3) sparse retrieval models \cite{splade,splade-v2,Choi2022SpaDEIS,doc2query}. The cross-encoder model is often parameterized with LLMs, such as BERT \cite{BERT} or T5 \cite{T5}, and takes the concatenation of query and document pair as input to predict their relevant score. This model is effective but slow and is usually used for re-ranking. As for retrieval, the dense retrieval model often uses the bi-encoder architecture to encode the query and document separately into the low-dimensional hidden space and apply the approximate nearest neighborhood (ANN) \cite{hnsw,ANN} search for fast retrieval.
%The techniques, such as negative sampling \cite{Adore,rocketqa,DPR,ANN}, knowledge-distillation \cite{tct-colbert,MarginMse,TAS-B,CL-DRD} and pre-training \cite{Gao2021UnsupervisedCA,ma2020bprop,Xiao2022RetroMAEPR} have been demonstrated useful for training an effective dense retrieval model. 
Sparse retrieval is an alternative method for retrieval, in which it encodes the query and document into the high-dimensional vector space, and usually, each element in the vector represents the importance score of a certain token. To filter out those useful tokens, the L1 \cite{snrm} or FLOPs \cite{splade,splade-v2,Paria2020MinimizingFT} regularizer will be incorporated into the objective function to sparsify the high-dimension vectors. For retrieval, the inverted index will be employed similar to BM25.  \\
\underline{\textbf{Generative Retrieval Models}}:
Generate Retrieval (GR), diverges from the traditional "index-then-retrieve" paradigm used in the sparse and dense retrieval models, offering a novel approach for document retrieval. In GR, each document is represented as a unique document identifier (DocID), and a sequence-to-sequence model is trained to generate relevant DocIDs given a query. 

DocIDs are usually fixed in the fine-tuning stage and hence serving as bottleneck for affecting the effectiveness of GR models. Usually, DocIDs fall into two categories: (1) semantic-based DocIDs, and (2) word-based DocIDs. Semantic-based DocIDs are usually created using quantization \cite{Ultron,tiger,RIPOR,Chen2023ContinualLF} or hierarchical clustering algorithms \cite{DSI,dsi++,NCI,Sun2023LearningTT} on document representations to capture semantic relationships among documents. 
%For instance, \citet{DSI} proposes DSI that employs the hierarchical k-means clustering on BERT-derived document representations to build a hierarchical tree. Then each document belongs to a root-to-leaf path which can be encoded as a sequence of integer numbers for a document identifier. 
In contrast, word-based DocIDs are directly constructed from the document content, including titles \cite{gere,corpus-brain,Lee2022NonparametricDF,Genre}, n-grams \cite{SEAL,MINDER,LTRGR,Wang2023NOVOLA,ugr}, URLs \cite{Ultron,Ren2023TOMEAT}, and significant words \cite{term-sets}.

During inference, search algorithms like constrained beam search  or FM-index are used to generate valid DocIDs given a query \cite{DSI,RIPOR,SEAL,MINDER}. As for fine-tuning, earlier studies \cite{DSI,dsi++,SEAL,NCI} directly optimize the model using the sequence-to-sequence cross-entropy loss. Recent research \cite{LTRGR,RIPOR} demonstrates that utilizing the learning-to-rank loss can further enhance the model performance. Data augmentation approaches, such as using pseudo queries \cite{Ultron,DSI-QG,NCI} are also proven to be useful as they can mitigate the distribution mismatches between the index and retrieval phases. While GR models have shown promising results on the small-scaled datasets, such as NQ-320K \cite{DSI} and MSMARCO-100K \cite{Pradeep2023HowDG}, their effectiveness in large-scale benchmarks remains a subject of debate \cite{Pradeep2023HowDG}. Recently, \citet{RIPOR} addressed this by introducing RIPOR--a framework that enhances the GR model with relevance-based DocID initialization and prefix-oriented ranking optimization. RIPOR has shown competitive performance to a number of strong dense retrieval methods on the standard MSMARCO data with 8.8M passages. 
 
\section{Methodology}
\subsection{Preliminaries and Motivations}
\subsubsection{\textbf{Generative Retrieval}}
%1) generative retrieval and its notation.
In generative retrieval, each document is symbolized with a unique identifier, which is commonly termed as \textit{DocID}. Generative retrieval models are often developed based on large language models to take a query string and generate a ranked list of DocIDs, with respect to their generation probability in descending order. Following the probability ranking principle \cite{robertson1977probability}, these generation probabilities are expected to model the probability of relevance for the corresponding documents. A constrained beam search algorithm \cite{DSI} is used for DocID decoding during inference. The decoded DocIDs are then mapped back to their corresponding documents, which form a final document ranking for the given query. 

Formally, let $M$ denote a generative model with an encoder-decoder architecture. The DocID for each document $d$ in the corpus $\mathcal{C}$ is represented as $c^d = [c_1^d, \ldots, c^d_L]$, where $L$ is the length of DocIDs. 
The model $M$ is often trained to generate the DocIDs autoregressively for any given query $q$. To generate the $i$\textsuperscript{th} DocID token $c^d_i$, the model is conditioned on the previously generated tokens, denoted as $c^d_{<i} = [c^d_1, \ldots c^d_{i-1}]$ as well as the query encoding. Therefore, the model generates the hidden representation for the DocID token $c^d_i$ as follows:
\begin{align}\label{eq:original_hidden_rep_each_step}
    \mathbf{h}_i^d = \text{Decoder} (\text{Encoder}(q), c^d_{<i}) \in \mathbb{R}^D 
\end{align}

Each DocID token is associated with a $D$-dimensional embedding. Let us assume $\mathbf{E}_i \in \mathbb{R}^{V \times D}$ represents the token embedding table at position $i$, where $V$ is the DocID vocabulary size.\footnote{Note that DocID vocabulary is different from the input vocabulary and may only contain some unique numbers.} Hence, the corresponding embedding for DocID token $c_i^d$ is represented as $\mathbf{E}_i[c^d_i] \in \mathbb{R}^D$. Note that, the embedding table at each position can be distinct, that is to say, $\mathbf{E}_i \neq \mathbf{E}_j: \forall i \neq j$.

We follow the scoring function introduced by RIPOR \cite{RIPOR}---the current state-of-the-art generative retrieval model---to compute the query-document relevance scores (i.e., the DocID generation score in response to the query) as follows: 
\begin{align}\label{eq:original_q_seq_docid_score}
    s(c^d; q) = \displaystyle\sum_{i=1}^L \mathbf{E}_i [c^d_i] \cdot \mathbf{h}^d_i
\end{align}

\subsubsection{\textbf{Constrained Beam Search}}
% The generative model $M$ generates the top k full-length DocIDs autoregressively from the start token in response to a given query $q$. 
The generative model $M$ often generates each DocID autoregressively using constrained beam search \cite{DSI,dsi++,RIPOR,NCI,DSI-QG}.
At each decoding step $i$, the beam search algorithm maintains the top $k$ prefixes with the highest probabilities (denoted by $P^{\text{topk}}_i$, where $|P^{\text{topk}}_i| = k$) %$\{ c^{j}_{< i}\}_{j=1}^k$ 
and expands each prefix by one token. Therefore, at each decoding step, many scored prefixes are pruned due to their low probability. The constrained beam search algorithm additionally uses a prefix tree \cite{DSI} to keep track of valid next tokens for each prefix. The prefix tree is built based on all DocIDs $\{ c^d: \forall d \in \mathcal{C}\}$, where $\mathcal{C}$ is the corpus. Therefore, constrained beam search ensures that every newly generated prefix belongs to at least one valid DocID. This can be accomplished using the following masking function $g$, defined for any sequence length $1 \leq i \leq L$, based on the prefix tree:
\begin{align*}
    g ([c_1, c_2, \cdots, c_i]) = \begin{cases}
        0 & \text{if } \ [c_1, c_2, \cdots, c_i] \ \text{is a valid prefix.} \\
        -\infty & \text{if } \ [c_1, c_2, \cdots, c_i] \ \text{is not a valid prefix.}
    \end{cases}
\end{align*}

% We assume that a new candidate token is termed as $c^x$ and is chosen from $\{1,  \ldots, V \}$, and a masking function $g$ based on the prefix tree is computed as:
% \begin{align*}
%     g (c^{j}_{<i}, c^x) = \begin{cases}
%         0 & \text{if } \ [c^{j}_{<i}, c^x] \ \text{is a valid prefix.} \\
%         -\infty & \text{if } \ [c^{j}_{<i}, c^x] \ \text{is not a valid prefix.}
%     \end{cases}
% \end{align*}

Therefore, at the $i$\textsuperscript{th} decoding step, the constrained beam search algorithm assigns the following score to expand each prefix $c^p_{<i} = [c_1, c_2, \cdots, c_{i-1}] \in P^{\text{topk}}_{i-1}$ by one token:
% $[c^{j}_{<i}, c^x] := c^{j,x}_{\leq i}$ as follows:
% \begin{align}\label{eq:beam_search}
%     s(c^{j,x}_{\leq i};q) &= s( c^{j}_{<i}; q) + s(c^x | q, c^{j}_{<i}) + g (c^{j}_{<i}, c^x) \nonumber \\ 
%                             &= s( c^{j}_{<i}; q) +  \mathbf{E}_i [c^x] \cdot \mathbf{h}_i^{j} + g (c^{j}_{<i}, c^x)
% \end{align}
\begin{align}\label{eq:beam_search}
    s(c^p_{\leq i}; q) &= s(c^p_{<i}; q) + s(c^p_{i} ; q, c^p_{<i}) + g (c^p_{\leq i}) \nonumber \\ 
        &= s(c^p_{<i}; q) +  \mathbf{E}_i [c^p_i] \cdot \mathbf{h}^p_i + g (c^p_{\leq i})
\end{align}

Based on the scoring function, the top $k$ expanded prefixes with highest probabilities are maintained for the next step. 

\subsubsection{\textbf{Pitfalls of (Constrained) Beam Search}}
\label{sec:motivation}
Beam search is a greedy local search algorithm that tends to get stuck into local optima instead of the global optimum \cite{Zhou2005BeamStackSI,Stahlberg2019OnNS}. \emph{Even though beam search has been successfully used in natural language generation \cite{Sutskever2014SequenceTS,Graves2012SequenceTW}, we hypothesize that it is not sufficient for developing effective generative retrieval models.} In language generation, there are multiple alternatives that can be equally acceptable outputs, e.g., grammatically correct sentences with same semantics. Hence, if a word token is dropped through beam search, it is likely that other equally good word tokens be kept for the next decoding step. However, in generative retrieval, each relevant document is represented with a unique DocID and if a prefix of this DocID is pruned by the beam search decoding algorithm, there is no way to recover, and that relevant document will not appear in the retrieval result list.

\begin{figure}
    \centering
    \includegraphics[width=.45\textwidth]{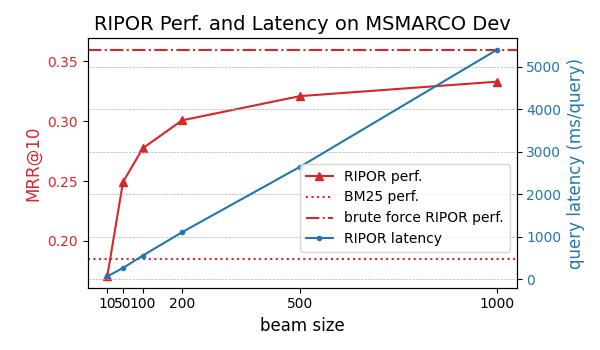}
    \caption{Retrieval effectiveness (MRR@10) and efficiency (query latency) of RIPOR \cite{RIPOR} w.r.t different beam sizes on the MS MARCO Dev Set -- a standard passage retrieval benchmark with 8.8M passages. The experiment is conducted on a single A100 GPU with 80GB memory. Best to be viewed in color.}
    \label{fig:RIPOR_perf_vs_bs}
    \vspace{-.5 cm}
\end{figure}

To empirically validate this hypothesis, we focused on the current state-of-the-art generative retrieval model, called RIPOR \cite{RIPOR}. RIPOR uses constrained beam search for DocID decoding. The efficiency and effectiveness of this model for various beam sizes on the MS MARCO Dev set \cite{MSMARCO} are plotted in \figurename~\ref{fig:RIPOR_perf_vs_bs}. We observe that increasing the beam size significantly impacts the effectiveness of RIPOR, even for a short list of 10 documents (MRR@10). Even with a beam size of 1000, RIPOR with constrained beam search cannot meet the brute force decoding performance. Here, brute force decoding means that every document in the corpus is scored by RIPOR without any prefix pruning. On the other hand, large beam size values lead to higher query latency, limiting the practicality of the models. These results validate our hypothesis that the prefixes of many relevant documents get harshly pruned by constrained beam search even with relatively large beam size values. This has motivated us to develop alternative decoding methods for generative retrieval models.

\begin{figure*}
    \centering
    \includegraphics{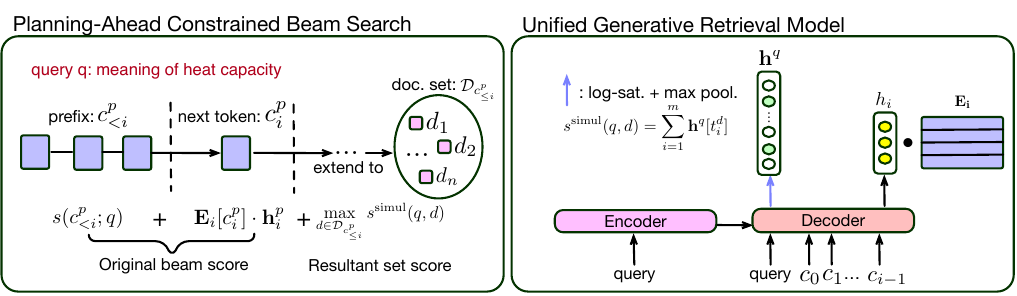}
    \caption{A visualization of the \framework framework. Left: Illustration of simultaneous decoding guiding autoregressive generation with approximate document-level scores. Right: illustration of the model $M$ employing joint decoding of set-based and sequential DocIDs.}
    \vspace{-.5 cm}
    \label{fig:GD-RIPOR-arch}
\end{figure*}

For this, we utilize the characteristic of the prefix tree and propose a novel approach that guides the autoregressive generation through simultaneous decoding, which the details will be elaborated in  \ref{sec:Globally-Guided-constrained-beam-search}. We create the set-based and sequential DocIDs to support the simultaneous and sequential decoding respectively, introduced in Section \ref{sec:DocID_construction_computation}. Additionally, we propose a three-stage training pipeline for gradual adaptation of the model to joint decoding, introduced in Section \ref{sec:optimization_details}. The high-level overview of the \framework framework is illustrated in Figure \ref{fig:GD-RIPOR-arch}.

%To fulfill the requirements, we construct a new type of document identifier termed as $w_d$ for each document $d$. We treat it as a bag-of-words representation, meaning that the relevant score between query $q$ and $w_d$ can be computed in a nonautogressive way (only call generative model once). To provide the complementary information with semantic-based DocIDs, 
%that is based on lexical method. We also view the new DocID as a bag-of-words, enabling the their relevant scores given a query can be computed fast by only calling the generative retrieval model once. We term this new DocID for each document $d$ as $w_d$.
%The detailed description on how to construct the new $w_d$ and original $c_d$ we be introduced in Section \ref{DocID_construction}.
%How $w_d$ is incorporated in generative retrieval model and optimized jointly with $c_d$ will be introduced in Section \ref{sec:optimization_details}.
%By utilizing the document-level (global) scores by $w_d$, we propose a novel \textit{xxx-guided constrained beam search} algorithm in \ref{bag-of-words_beam_search}.
%We will explain how construct the new type of DocID in Section \ref{DocID_construction}. How are they incorporated into the RIPOR, and optimized jointly with original DocIDs in \ref{sec:optimization_details}. By utilizing the %document-level relevant scores from these new DocIDs,

\subsection{Planning-Ahead Constrained Beam Search}\label{sec:Globally-Guided-constrained-beam-search}
%The characteristic of \textit{constrained} beam search ensures that every expanded prefix $c^{j,x}_{\leq i}$ at step $i$ will extend to a set of valid DocIDs and then mapped to their corresponding documents upon decoding finished. Instead of scoring the expanded prefix solely based on the next token score used in Equation \ref{eq:beam_search}, this characteristic gives us great flexibility to incorporate the global score derived from the resultant document set into the original local beam search. The global score would up-weight the prefixes with the final sets that contain relevant documents and down-weight those irrelevant ones. The re-weighting mechanism increases the survival chance of relevant prefixes and mitigates the issue of local beam search.

The prefix tree used in the constrained beam search ensures that every expanded prefix  $c^{p}_{\leq i}$ at step $i$ is a valid sequence, thus leading to a set of valid DocIDs once decoding finishes. However, according to Equation~\eqref{eq:beam_search}, which is used in state-of-the-art generative retrieval models, the document ID prefixes are expanded solely based on the contribution by the next token score. Our motivative experiments in Section~\ref{sec:motivation}, on the other hand, demonstrates that this is not sufficient and the prefixes of many relevant documents get pruned in constrained beam search, even with large beam size values. This section introduces a novel approach for generative retrieval by planning sequential decoding using an efficient simultaneous decoding that approximates document-level scores. These scores are considered as priors for sequential decoding and let the decoding phase keep the tokens that are likely to lead to high relevance scores once decoding is finished.

\subsubsection{\textbf{Simultaneous Decoding}} 
Here we introduce an approach called \emph{simultaneous decoding} that produces a score for each document in one decoding step, using $s^{\text{simul}} (q, d)$. The next subsection incorporates this simultaneous document-level decoding into sequential decoding of autoregressive models.  
% As for the design to compute the document-level relevant score , we require it can be seamlessly incorporated into the generative retrieval model $M$ to facilitate the end-to-end training and also be computed fast that does not introduce too much overhead on the original autoregressive decoding process. 
To this aim, for each document $d \in \mathcal{C}$, we construct a new type of set-based DocIDs $t^d$, consisting of a \emph{set} of tokens $\{t^d_1, t^d_2, \cdots, t^d_{m}\}$, where $m$ is the set size for each document $d$. Note that $m$ is a constant for all documents and is a hyper-parameter. Unlike $c^d$, $t^d$ is a set and there is no particular order for tokens in $t^d$, hence they can be decoded simultaneously.

To compute the simultaneous decoding scores for a given query $q$, we feed the query text to the generative model $M$ to obtain the contextualized representations: $\mathbf{Q} = \text{Decoder} (\text{Encoder} (q), q ) \in \mathbb{R}^{|q| \times D}$, where $|q|$ and $D$ respectively represent the query length and the output embedding dimensionality for each token. Let $V_T$ denote the vocabulary size for DocIDs in simultaneous decoding, and $\mathbf{E}_{\text{simul}} \in \mathbb{R}^{V_T \times D}$ denote the associated embedding matrix. Inspired by \citet{splade,splade-v2}, we apply log-saturation and max pooling operations to compute a weight per DocID token. 
\begin{align}\label{eq:sparse_query_reps}
    \mathbf{h}^q = \text{MaxPool} \big (  \log (1 + \text{Relu} (\mathbf{E}_{\text{simul}} \cdot \mathbf{Q}^T ) ) \big) \in \mathbb{R}^{V_{T}}
\end{align}

Then the document-level simultaneous relevance score for every $(q, d)$ pair is then computed as:
\begin{align}\label{eq:global_score}
    s^{\text{simul}}(q, d) =  \displaystyle\sum_{i=1}^m \mathbf{h}^q[t^d_i]
\end{align}
$s^{\text{simul}}(q, d)$ can also be written as $s^{\text{simul}}(q, t^d)$.

\subsubsection{\textbf{Guiding Autoregressive Generation through Simultaneous Decoding}}
\framework uses simultaneous document scoring as priors for computing prefix scores in autoregressive generation. In other words, for decoding any prefix, we consider the maximum approximate document score that can be associated with that prefix as prior. There exist other aggregation functions, such as $\text{mean} (\cdot)$ or $\min (\cdot)$, to consume here, however, we empirically found $\max (\cdot)$ superior. Formally, let $\mathcal{D}$ be a set of top $n$ documents with the highest scores, according to Equation~\eqref{eq:global_score}. 
We rewrite Equation~\eqref{eq:beam_search} as:
\begin{align}\label{eq:scoring_function}
    s'(c^p_{\leq i};q) &= \max_{d \in \mathcal{D}_{c^p_{\leq i}}} s^{\text{simul}} (q, d)  + s(c^p_{\leq i};q)   \\
    &= \max_{d \in \mathcal{D}_{c^p_{\leq i}}} s^{\text{simul}} (q, d)  + s( c^p_{<i}; q) +  \mathbf{E}_i [c^p_i] \cdot \mathbf{h}_i^{p} + g(c^p_{\leq i}) \nonumber
\end{align}
where $\mathcal{D}_{c^p_{\leq i}} = \{d \in \mathcal{D} | c^p_{\leq i} = c^d_{\leq i}\}$ is a subset of all documents from $\mathcal{D}$ with the prefix of $c^p_{\leq i}$. 
This modified scoring function conditions the next token decoding on an approximate of (future) resultant document score through simultaneous scoring. This can minimize the impact of aggressive document ID pruning in the original beam search algorithm. Based on the modified prefix scoring function, we propose a novel decoding method for generative retrieval and term it as \emph{planning-ahead constrained beam search}. This decoding process is illustrated in Algorithm \ref{alg:modified_beam_search}.
\vspace{-.15cm}
\subsubsection{\textbf{Computational Cost of Decoding}}
To assess the computational costs of sequential and simultaneous decoding, we utilize Floating Point Operations (FLOPs). The sequential decoding mainly incurs costs from multiple forward calls of the generative model $M$. Assuming a beam size of $k$, a sequential DocID length $L$, and $P_m$ FLOPs per forward call of $M$, the total FLOPs for sequential decoding are approximately $P_{\text{seq}} = L \cdot k \cdot P_m$. In contrast, simultaneous decoding involves a single forward call of $M$ and additional computations for generating simultaneous relevance scores across the corpus using Equation \eqref{eq:global_score}. If the corpus size is $|\mathcal{C}|$, the total FLOPs for simultaneous decoding is $P_{\text{simul}} = P_m + |\mathcal{C}| \cdot m$. The FLOPs difference is $\Delta P = P_{\text{seq}} - P_{\text{simul}} = P_m \cdot (L \cdot k - 1) - \mathcal{C} \cdot m$. 

Let us assume $M$ is T5-base, a language model that is often studied for generative retrieval \cite{RIPOR,dsi++,DSI,NCI} and is considered relatively small compared to today's landscape of LLMs. It requires approximately $7.5\times 10^{9}$ FLOPs per forward call and given the size of retrieval collections like MSMARCO's 8.8 million passages (used in this study), we infer that simultaneous decoding is notably faster than sequential decoding. This is empirically supported by the query latency comparison in Table \ref{tab:bow_and_beam_size} in our experiments. While our focus in this paper is on the million-scale dataset, it is important to note that for billion-scale collections, the simultaneous decoding, if done brute-force, might be slower. Approximation techniques like \cite{Jgou2011SearchingIO,Babenko2012TheIM,Baranchuk2018RevisitingTI,Johnson2017BillionScaleSS} could be used for maintaining simultaneous decoding's efficiency at billion-scale. We acknowledge that further exploration is needed in the future.

\begin{algorithm}
\caption{Planning-Ahead Constrained Beam Search}
\label{alg:modified_beam_search}
\begin{algorithmic}[1]
\REQUIRE Generative retrieval model $M$, query $q$, beam size $k$, retrieval corpus $\mathcal{C}$, set-based DocIDs $\{ t^d \}_{d \in \mathcal{C}}$, sequential DocIDs $\{ c^d \}_{d \in \mathcal{C}}$, vocabulary size for each token embedding $V$, sequential DocID length $L$.

%\STATE Pre-compute document-level scores for every $t^d$: $s^{\text{simul}} (q, t^d)$ using Eq. \eqref{eq:global_score}.
\STATE Pre-compute document-level scores for every $t^d$: $s^{\text{simul}} (q, t^d)$ using Eq. \eqref{eq:global_score}, and select the top $n$ documents to construct the set $\mathcal{D}$.
\STATE Find every possible prefix $c^{*}_{\leq i}$ and the resultant set $\mathcal{D}_{c^*_{\leq i}}$ from $\mathcal{D}$. Based on that, construct a dictionary $T$, where the key is  $c^{*}_{\leq i}$ and the value is $\displaystyle\max_{d \in \mathcal{D}_{c^{*}_{\leq i}}}  s^{\text{simul}} (q,t^d)$. 
% \hansi{Please check whether the writing convinces you. If I write the complete process for dictionary $T$ construction, I feel like it is too long.} \hamed{good for now. we can think more about it for camera-ready.}
\STATE  $B_1 \gets \{ <0,0,\text{BOS}> \}$
\FOR{$i = 1$ to $L$}
\STATE $B \gets \emptyset$
    \FOR{ $(s,s',c^p_{ < i}) \in B_i$ }
        \FOR {$c^p_i \in V$}
        \STATE $c^p_{\leq i} \gets [c^p_{<i}, c^p_i]$, $\displaystyle\max_{d \in \mathcal{D}_{c^{p}_{\leq i}}}  s^{\text{simul}} (q,t^d) \gets T[c^p_{\leq i}]$ 
        \STATE Apply Eq. (\ref{eq:scoring_function}) and $B$.add\big($<s(c^p_{\leq i};q)$, $s'(c^p_{\leq i};q)$, $c^p_{\leq i}>$\big)
        \ENDFOR
    \ENDFOR
\STATE $B_{i+1} \gets B.\text{top} (k)$ based on the $s'(\cdot)$. (the second element)
\ENDFOR
\RETURN $B_{L+1}$ (the third element is DocID, and the second element is corresponding relevant score)

\end{algorithmic}
\end{algorithm}

%Assume we obtain sequence based DocID $c_d = \{ c^d_1, \ldots c^d_L \}$ and set based DocID $w_d = \{ w^d_1, \ldots w^d_U \}$ for $\forall d \in \mathcal{C}$. 

%We elaborate the Globally-Guided constrained beam search in this section. After DocID construction process, we can obtain the sequence based DocID $c_d = \{ c^d_1, \ldots c^d_L \}$ and set based DocID $w_d = \{ w^d_1, \ldots w^d_U \}$ for $\forall d \in \mathcal{C}$. Given a query $q$, the generative retrieval model $M$ is first to generate the global score for every $w_d$. Specifically, a query $q$ will be taken as input into the encoder and decoder of model $M$, and then go through the log-saturation and max pooling layerss, similar to Equation \ref{eq:sparse_reps}:
%\begin{align*}
%    \mathbf{h}_q^{sp} = \text{MaxPool} \big (  1 + \text{Relu} (\mathbf{E}_{\text{T5}} \cdot \mathbf{H}_q^T )  \big) \in \mathbb{R}^{V_{T5}}
%\end{align*}
%Then the global score for every $w_d$ given $q$ is:
%\begin{align}\label{eq:global_score}
%    s^{\text{global}}(q, w_d) =  \displaystyle\sum_{i=1}^U \mathbf{h}_q^{sp}[w^d_i]
%\end{align}
%Since the model $M$ is only called once for the query $q$, the scoring process over the all $w_d$ could be completed quickly. The global scores will be pre-stored and used for constrained beam search.

\vspace{-.35 cm}
\subsection{DocID Construction}\label{sec:DocID_construction_computation}

We can envision multiple approaches for constructing the sequence- and the set-based document identifiers. Without loss of generality, in the following, we describe the approach we used in this paper.

\subsubsection{\textbf{Sequential DocID Construction}}\label{sec:sequence_based_docid_construction}
We follow the approach of relevance-based DocID initialization introduced in RIPOR \cite{RIPOR} to construct the sequential DocIDs (i.e., $c^d$s), in which we treat the generative model $M$ as a dense encoder. We utilize the encoder-decoder architecture of $M$ by feeding a document text to the encoder and a start token to the decoder. The document representation is then obtained by the contextualized output representations of the decoder: 
\begin{align}\label{eq:dense_rep}
    \mathbf{d} = \text{Decoder} (s_0, \text{Encoder} (d)) \in \mathbb{R}^D
\end{align}
where $s_0$ is the start token. By using the $M$ as the dense encoder, we can obtain the dense representation $\mathbf{d}$ for each document $d$. Subsequently, employing the residual quantization (RQ) algorithm \cite{RQ}, we compile a token embedding tables $\{ \mathbf{E}_i \}_{i=1}^L$ to determine the DocID $c^d = [c_1^d, \ldots c_L^d] $ for each document $d \in \mathcal{C}$. This optimization process would make each representation $\mathbf{d}$ be approximated as the sequence of token embeddings:  
\begin{align*}
    \mathbf{d} \approx \displaystyle\sum_{i=1}^L \mathbf{E}_i [c^d_i]
\end{align*}

\subsubsection{\textbf{Set-Based DocID Construction}}
The sequential DocID $c^d$ captures the document's semantic information by applying the RQ on derived dense representation $d$. In the realm of IR, it is well-acknowledged that combining the semantic and lexical information of documents would enhance the retrieval system performance \cite{Chen2021SalientPA,tct-colbert,Wang2021BERTbasedDR,Lin2021AFB,Zhang2022LEDLD}. With this motivation, we set $V_T$ to the vocabulary size of our generative model $M$, and $\mathbf{E}_\text{simul}$ to the embedding table in $M$.  Hence, the tokens $\{t_1^d, t_2^d, \ldots t_m^d\}$ for each set-based DocID $t^d$ will be directly constructed based on the tokenized document content. There exist various methods to score and extract the representative tokens from documents \cite{SprckJones2021ASI,Lin2021AFB,splade-v2,splade}. For the scoring part, the traditional methods, such as TF-IDF \cite{SprckJones2021ASI}, weight each token using the corresponding term statistics, e.g., term frequency and inverse document frequency. With the recent advancement of neural lexical models \cite{splade-v2}, the term importance scores can be directly learned from the supervised training data. 

% We adopt Splade \cite{splade-v2} to learn to weight the sub-words in the T5-vocabulary for each document. 
Similar to Equation~\eqref{eq:sparse_query_reps}, we take the document content $d$ as the input to the encoder and decoder of the model $M$ to obtain the contextual representation $\mathbf{h}^d \in \mathbb{R}^{V_{T}}$.
Then we follow the training objective used in \cite{splade-v2} by linearly combining the MargiMSE loss (retrieval-oriented objective) with FLOPs regularizer \cite{splade,flop} to sparsify  document  representations. We describe the training process in Section \ref{sec:optimization_details} in detail.
The non-zero weights in $\mathbf{h}^d$ represent the importance score of the corresponding tokens. Hence, we select the top $m$ tokens to form the set-based DocID $t^d = \{t_1^d, \ldots t_m^d\}$ for each document $d$. 
%We adopt Splade \cite{} to learn to weight the sub-words in T5-vocabulary for each document. Splade applies the log-saturation and max pooling operation over the Masked Language Modeling (MLM) layer on contextualized representations on a given document content $d$ to obtain the corresponding high-dimensional representation. Specifically, we take the document content as the input to the encoder and decoder of generative model to obtain the contextualized representation:
%\begin{align*}
%    \mathbf{H}_d = \text{Decoder} (\text{Encoder} (d), d ) \in \mathbb{R}^{L_d \times D} 
%\end{align*}
%Where $L_d$ is the length of tokenized document content $d$. Then the high-dimensional representation for document $d$ is derived from the following equation:
%\begin{align}\label{eq:sparse_reps}
%    \mathbf{h}_d^{sp} = \text{MaxPool} \big (  \log (1 + \text{Relu} (\mathbf{E}_{\text{T5}} \cdot \mathbf{H}_d^T ) ) \big) \in \mathbb{R}^{V_{T5}}
%\end{align}
%Where $V_{T5}$ is the T5 vocabulary size. Splade will apply FLOPs regularizer \cite{} and retrieval-oriented objective, such as MarginMSE \cite{} to obtain the sparse representation for this high-dimensional vector. We will describe the training process in Sectin \ref{sec:optimization_details}. 
%For each of the sparse representation $\mathbf{h}_d^{sp}$, the non-zero weights represent the importance score for the corresponding sub-words in the T5-vocabulary. We select the top U sub-words to form the set-based DocID $w_d = \{w_1^d, \ldots w_U^d\}$ for each document $d$. 

\subsection{\framework Optimization}\label{sec:optimization_details}
The whole optimization process consists of three stages, the first two of which can be trained in parallel. The first two stages are applied to make generative retrieval model capable of predicting set-based DocIDs and sequential DocIDs, respectively. In the final stage, we jointly train the two types of DocIDs together in a unified model, which makes it suitable for the planning-ahead constrained beam search decoding introduced in Section \ref{sec:Globally-Guided-constrained-beam-search}.

\subsubsection{\textbf{Generative Retrieval Model for Set-based DocIDs}}: This stage contains two training phases: (1) the pre-training phase is used for obtaining the set-based DocIDs and model warmup; (2) the fine-tuning phase is used to train the generative retrieval model for set-based DocIDs prediction. 

\underline{\textbf{Pre-training}}:
To obtain the set-based DocIDs $t^d$, we first treat the generative retrieval model $M$ as a sparse encoder. Given any triple $(q, d^+, d^-)$, where $d^+$ and $d^-$ represent a relevant and a non-relevant document for $q$. We follow the MarginMSE method \cite{MarginMse} to obtain the teacher margin $T(q, d^+, d^-) = S^T(q, d^+) - S^T(q, d^-)$ for each triplet. Usually, the teacher relevance scores $S^T(q, d)$ for each $(q,d)$ pair is derived from the cross-encoder based teacher model \cite{MarginMse,TAS-B}. We obtain the sparse representations for $q$, $d^+$, $d^-$ using Equation \eqref{eq:sparse_query_reps}. Based on the sparse representations, we apply the training objective of \citet{splade-v2} with MarginMSE loss and FLOPs regularizer \cite{flop}. After pre-training, we can apply the ``sparse encoder'' to obtain the sparse representation $\mathbf{h}^d$ for every document $d$. Each non-zero element in the sparse vector represents the important score for the corresponding token. We select top $m$ tokens for each document $d$ as the set-based DocID $t^d$. We denote the trained model as $M^{sp}$.

\underline{\textbf{Fine-tuning}}:
We first use $M^{sp}$ as the negative sampler to retrieve the top 100 documents for each query $q$ and denote the set as $\mathcal{D}^{sp}_{q}$. We construct the training triples: $(q, t^{d^+}, t^{d^-})$, $d^- \in \mathcal{D}^{sp}_{q}$ for fine-tuning the generative retrieval model. The model is initialized with $M^{sp}$. We use Equation \eqref{eq:global_score} to compute the relevance score for each $(q, d)$ pair. The loss function for each triplet is computed as:
\begin{align*}
    \mathcal{L}_{\text{set}} (q, t^{d^+}, t^{d^-}) = \big(s^{\text{simul}} (q, t^{d^+})- s^{\text{simul}} (q, t^{d^-}) - T(q, d^+, d^-) \big)^2 
\end{align*}
Where $T(\cdot)$ is the teacher margin the same as in the pre-training stage. Motivated by the self-negative training strategy \cite{ANN,rocketqa,Adore}, we use the trained model as negative sampler to sample top 100 documents, and merge the negative document set with $\mathcal{D}^{sp}_{q}$ for each query $q$, then we apply the same $ \mathcal{L}_{\text{set}}$ training loss to fine-tune the model again, and we denote the trained model as $M^{\text{set}}$.

\subsubsection{\textbf{Generative Retrieval Model for sequential DocIDs}}
Similarly, this training stage contains two phases. The first phase is to construct the sequential DocIDs and the second phase is to fine-tune the generative retrieval model. 

\underline{\textbf{Pre-training}}:
The goal of this stage is to obtain the sequential DocIDs $c^d$. We use the same method as RIPOR \cite{RIPOR} that treats the generative retrieval model as a dense encoder and applies the residual quantization (RQ) on the obtained dense representations. We apply the MarginMSE loss with a two-step training strategy to train the model. To construct the training triplets, the negative documents in the first step are sampled from the top 100 documents of BM25 and the negatives in the second step are sampled from the top 100 of trained model itself at first step. As introduced in Section \ref{sec:sequence_based_docid_construction}, we obtain the dense representation for each document $d$ by using Equation \eqref{eq:dense_rep}, and then applying RQ to obtain the sequential DocIDs. We obtain the trained model $M^{ds}$, and the corresponding token embeddings $\{ \mathbf{E}_i \}_{i=1}^L$ after the dense retrieval pre-training.

Similar to RIPOR, this stage also contains a seq2seq pre-training phase. Specifically, we use the doc2query model \cite{doc2query} to generate 10 pseudo-queries for each document, then we take each of the pseudo-queries as input to model and predict the corresponding sequential DocID using a seq2seq loss. The model is initialized with $M^{ds}$ and we denoted the trained model as $M^{s2s}$.

\underline{\textbf{Fine-tuning}}:
Similar to the previous fine-tuning stage, we construct the training triplets by using the $M^{ds}$ to sample top 100 negative documents and denote the negative set as $\mathcal{D}^{ds}_q$ for each query $q$. We apply the multi-objective training loss in RIPOR \cite{RIPOR} for prefix-oriented fine-tuning. The full-length relevance score between $(q, c^d)$ can be computed via Equation \eqref{eq:original_q_seq_docid_score}. 
Similarly, the relevance score by generative retrieval model produced by the first $i$ tokens of $c^d$: $[c^d_1, \ldots c^d_i ]$ can be computed via Equation \eqref{eq:beam_search}. Given any triplet $(q, c^{d^+}, c^{d^-} )$, we can use the modified MarginMSE loss for each prefix with length $i$:
\begin{align*}
    \mathcal{L}_{\text{seq}}^i =\big( s(c^{d^+}_{\leq i}; q) - 
    s(c^{d^-}_{\leq i}; q) - 
    \alpha_i T(q, d^+, d^-) \big)^2
\end{align*}
where $\alpha_i$ is the monotonically increasing weight w.r.t $i$ ranging from [0, 1], and $\alpha_L = 1$. Refer to \citet{RIPOR} for the details. Therefore, the multi-objective loss is:
\begin{align*}
    \mathcal{L}_{\text{seq}} = \displaystyle\sum_{i \in S_L} \mathcal{L}_{\text{seq}}^i
\end{align*}
$S_L$ is a set containing the sampled prefix lengths and $L$. The optimized model is termed as $M^{\text{seq}}$ which starts from $M^{s2s}$. Once $M^{\text{seq}}$ is trained, we use it as the negative sampler for sampling top 100 documents for each query $q$, denoted as $\mathcal{D}^{sq}$.

\subsubsection{\textbf{Unified Optimization for Generative Retrieval with Set-based \& Sequential DocIDs}}
In this stage, we train a single model that can predict the set-based DocIDs and sequential DocIDs jointly, which makes it capable of the proposed \emph{planning-ahead constrained beam search}. We initialize the weights of the generative retrieval model $M$ by averaging the weights of $M^{\text{set}}$ and $M^{\text{seq}}$. We use the two types of DocIDs to construct the training triples: $\{ (q, t^{d^+}, t^{d^-}), (q, c^{d^+}, c^{d^-}) \}$ and the negative sample set is $\mathcal{D}^{sp}_q \cup \mathcal{D}^{sq}_q$ for each query $q$. We obtain the score $s^{\text{simul}}(q,d)$ as Equation \eqref{eq:global_score} for each $(q, d)$ pair. To be compatible with the document-level simultaneous relevance score computation, we apply the slight modification for model $M$ to generate the hidden representation for the next token $c_i^d$. Different from Equation \eqref{eq:original_hidden_rep_each_step}, we additionally feed the query content to the decoder: 
\begin{align*}
    \mathbf{h}_i^d = \text{Decoder} ( \text{Encoder} (q), q, c^d_{<i}) \in \mathbb{R}^D
\end{align*}
Based on this, $s(c^d;q)$ is computed the same as Equation \eqref{eq:original_q_seq_docid_score}. Therefore, we use the following objective to train the unified generative retrieval model for each triplet:
\begin{align*}
    \mathcal{L}_\text{set} (q, t^{d^+}, t^{d^-}) + \mathcal{L}_{\text{seq}} (q, c^{d^+}, c^{d^-})
\end{align*}

%\hamed{overall comment about the optimization section: It is well-written and detailed, but it can be confusing as it has too many details. Make sure to read it once thoroughly and see if you can simplify things.}

\section{Experiments}
\begin{table*}
    \centering
    \caption{Experimental results on MSMARCO and TREC Deep Learning Track Data. Highest generative retrieval performances are boldfaced. Superscript $*$ denotes statistically significant improvement compared to all generative retrieval baselines. Superscripts $^\triangle$ and $^\triangledown$ denote significantly higher and lower performance compared to \framework for sparse and dense retrieval models. (t-test with Bonferroni correction, p\_value < 0.01). We use brute-force search for dense retrieval models.}
    \begin{tabular}
    {p{3cm}ll!{\color{lightgray}\vrule}ll!{\color{lightgray}\vrule}ll!
    {\color{lightgray}\vrule}ll!}
    \toprule
    % \multirow{3}{*}{\textbf{Model}} & \multirow{3}{*}{\parbox{1cm}{\textbf{Index} \\ 
    \multirow{2}{*}{\textbf{Model}} &  \multirow{2}{*}{\textbf{KD}} & \multirow{2}{*}{\parbox{1.5cm}{ \textbf{Index} \\ \textbf{Mem.(GB)}}}  & \multicolumn{2}{c!{\color{lightgray}\vrule}}{\textbf{MSMARCO Dev}} & \multicolumn{2}{c!{\color{lightgray}\vrule}}{\textbf{TREC DL 2019}} & \multicolumn{2}{c!}{\textbf{TREC DL 2020}} 
    \\
    & &  &  MRR@10 & Recall@10 & NDCG@10 & Recall@10 & NDCG@10 & Recall@10 \\
     \midrule
     \multicolumn{4}{l}{\textbf{Generative Retrieval Methods}} \\
     DSI  & \xmark & \phantom{0}0.03 & .045 & .138 & .163 & .076 & .150 & .070  \\
     DSI-QG  & \xmark & \phantom{0}0.03 & .105 & .292 & .320 & .138 & .328 & .120  \\
     Ultron-PQ & \xmark & \phantom{0}0.84 & .117 & .248 & .331 & .135 & .342 & .134 \\
     NCI-QG & \xmark & \phantom{0}0.03 & .153 & .352 & .403 & .167 & .394 & .159  \\
     SEAL & \xmark & \phantom{0}\phantom{0}- & .127 & -  & - & - & - & -  \\ 
     NOVO & \xmark & \phantom{0}0.80 & .126 & .242 & .258 & .112 & .310 & .140 \\ 
     MINDER  & \xmark &  12.16 & .186 & .383 & .506 & .201 & .392 & .144 \\
     LTRGR  & \xmark & 12.16 & .255 & .531  & .598 & .238  & .553 & .182 \\
     RIPOR  & \checkmark & \phantom{0}1.06 & .333 & .562 & .628 & .205 & .631 & .191  \\
     \framework  & \checkmark & \phantom{0}3.27 & \textbf{.385}$^*$ & \textbf{.670}$^*$ & \textbf{.705}$^*$ & \textbf{.267}$^*$ & \textbf{.700}$^*$ & \textbf{.236}$^*$ \\
     \midrule
     \multicolumn{4}{l}{\textbf{Some Sparse and Dense Retrieval Methods (For Reference)}} \\
     BM25 & \xmark &  \phantom{0}4.00 &  .185$^\triangledown$ & .381$^\triangledown$  & .512$^\triangledown$ & .178$^\triangledown$  & .477$^\triangledown$ & .164$^\triangledown$ \\
      docT5query & \xmark & 13.00 & .272$^\triangledown$ & .536$^\triangledown$  & .642$^\triangledown$ & .247$^\triangledown$  & .619$^\triangledown$  & .224$^\triangledown$  \\
      % Splade-v2 & \checkmark & - & .368 & - & .723 & - & - & - \\
     % \midrule
     % \multicolumn{4}{l}{\textbf{Dense Retrieval (For Reference)}} \\
     % \multicolumn{4}{l}{\textbf{Dense Retrieval (For Reference)}} \\
     %DPR & \xmark & 25.30  & .287$^\triangledown$ & .539$^\triangledown$ & .588$^\triangledown$ & .195$^\triangledown$  & .581$^\triangledown$ & .182$^\triangledown$ \\
     ANCE & \xmark & 25.30   & .330$^\triangledown$ &  566$^\triangledown$ &  .648$^\triangledown$ & .239$^\triangledown$ & .646$^\triangledown$ & .185$^\triangledown$  \\
     ADORE & \xmark & 25.30 & .347$^\triangledown$ & .611$^\triangledown$ & .683$^\triangledown$ & .264$^\triangledown$ & .666$^\triangledown$ & .214$^\triangledown$ \\
     RocketQA   & \checkmark & 25.30  & .370 & -  & - & -  & - & -   \\
    TCT-ColBERT  & \checkmark & 25.30  & .335$^\triangledown$ & .596$^\triangledown$  & .670$^\triangledown$ & .240$^\triangledown$  & .668$^\triangledown$ &  .218$^\triangledown$   \\
     MarginMSE  & \checkmark & 25.30  & .325$^\triangledown$ & .581$^\triangledown$ & .699$^\triangledown$ & .250$^\triangledown$ & .645$^\triangledown$ & .203$^\triangledown$ \\
     TAS-B  & \checkmark & 25.30  & .344$^\triangledown$ & .622$^\triangledown$  & .717$^\triangle$ & .255$^\triangledown$ & .685$^\triangledown$ & .230$^\triangledown$ \\
     CL-DRD & \checkmark & 25.30  & .382$^\triangledown$ & .651$^\triangledown$ & .725$^\triangle$ & .266  & .687$^\triangledown$ & .216$^\triangledown$ \\
     \bottomrule
    \end{tabular}
    \label{table:main_result}
\end{table*}

\subsection{Experiment Settings}
\subsubsection{\textbf{Datasets}}
We assess our model on the MSMARCO passage retrieval benchmark \cite{MSMARCO} comprising 8.8M passages and 532K train queries. Each trained query contains ~1.1 relevant passages on average. We evaluate our model using three evaluation datasets: (1) MSMARCO Dev with 6980 queries with incomplete relevance annotations; (2, 3) TREC-DL 2019 \& 2020: the passage retrieval datasets are used in the first and second iterations of TREC Deep Learning Track \cite{Trec-19,Trec-20}, which contains 43 and 54 queries respectively with a relatively complete relevance annotations done by TREC via pooling. For evaluation, we use the official metric for each dataset: (1) MRR@10 for MSMARCO Dev; (2) NDCG@10 for TREC-DL 19 and 20. We additionally follow \citet{RIPOR} and use Recall@10 for all the three datasets. 

\subsubsection{\textbf{Implementation Details}}
We apply T5-base \cite{T5} as the backbone for our generative retrieval model $M$. For sequential DocID initialization, we apply the residual quantization (RQ) by Faiss \cite{faiss} implementation. The sequential DocID length $L$ is set to $8$ and the vocabulary size $V$ is set to $2048$. As for hyper-parameters used in set-based DocIDs, the size of selected tokens ($m$) is $64$. 
In the pre-training and fine-tuning stages for set-based DocIDs, we set the learning rate to $0.0005$ and training epochs to $100$. The weights for the FLOPs regularization \cite{flop,splade-v2,splade} for query and document representations are $0.01$ and $0.008$, respectively.
In the dense encoder pre-training and prefix-oriented fine-tuning stages for sequential DocIDs, the learning rate is also set to  $0.0005$ and the training epochs are $50$ and $150$, respectively. The $S_L = \{4, 8\}$, and the corresponding weights $\alpha_i$s are $\{0.5, 1.0 \}$ In the seq2seq pre-training stage, the learning rate is $0.001$ and the number of training steps is $250,000$. 
For the final unified training stage, the learning rate is $0.0005$ and the number of training epochs is set to $120$. We use Adam optimizer \cite{Kingma2014AdamAM} with the linear warmup scheduling, the warmup ratio is set to $4.5\%$ of the total training steps. 
The beam size is $100$, and the top $1000$ documents are selected to form $\mathcal{D}$ for inference.
The model is trained on 8 A100 GPUs each with 40GB memory. For fair comparison in terms of efficiency, we use an A100 GPU with 80GB memory for all models at inference.
\vspace{-.25 cm}
\subsubsection{\textbf{Baselines}}
We compare our model with the following generative retrieval models: DSI \cite{DSI}, DSI-QG \cite{DSI-QG}, NCI-QG \cite{NCI}, Utron-PQ \cite{Ultron}, SEAL \cite{SEAL}, NOVO \cite{Wang2023NOVOLA}, MINDER \cite{MINDER}, LTRGR \cite{LTRGR} and RIPOR \cite{RIPOR}. To the best of our knowledge, RIPOR provides the strongest performance among all generative retrieval baselines.  
We also select the following competitive sparse and dense retrieval methods as points of reference: BM25 \cite{BM25}, docT5query \cite{doc2query}, ANCE \cite{ANN}, ADORE \cite{Adore}, RocketQA \cite{rocketqa}, TCT-ColBERT \cite{tct-colbert}, MarginMSE \cite{MarginMse}, TAS-B \cite{TAS-B}, and CL-DRD \cite{CL-DRD}. 

% Note that we do not claim improvements over all possible retrieval models. The goal of this paper is to introduce a novel optimization and decoding technique for generative retrieval that significantly advances the state-of-the-art in generative retrieval.

\subsubsection{\textbf{Comparison with Baselines}}
The comparison between baselines and \framework is illustrated in Table \ref{table:main_result}. First, \framework consistently outperforms other generative retrieval baselines across the three datasets. Compared with RIPOR that also uses knowledge distillation, \framework achieves $15.6\%$ relative improvement on MRR@10 in MSMARCO Dev set, which emphasizes the importance of adding set-based DocIDs constructed by the lexical approach for computing relevance scores. Notice that the beam size used in \framework is $100$, which is 10 times smaller than the beam size of $1000$ used in RIPOR. This can reduce the query latency significantly and implies the effectiveness of employing planning-ahead constrained beam search. 
Second, compared with the dense retrieval methods in our experiments, \framework also consistently shows better performance while using less index memory. For instance, \framework improves MRR@10 by $11.9\%$ over TAS-B, and in the TREC-DL 20 set, it leads to $2.2\%$ improvement on NDCG@10. It is also important to note that \framework uses $\times 7.7$ less index memory compared with dense retrieval models. %This reduced memory footprint might potentially offer a great advantage when deploying retrieval models in the billion-scale dataset. 
%Third, we observe that knowledge distillation is a useful technique to mitigate the incomplete label annotation issue in the MSMARCO training set. For example, in the generative retrieval category, LTRGR lags behind RIPOR and \framework by large margins. \hamed{but KD is not their only difference. Maybe you want to remove this claim.} \hansi{How about deleting point three if it is not clear?} \hamed{yes, do it} As for dense retrieval baselines, ADORE, the best-performing model without using teacher signals, consistently underperforms those knowledge distillation based models, such as TAS-B.
\vspace{-0.35cm}
\subsubsection{\textbf{Ablation Studies}}
We conduct a thorough ablation study on MSMARCO dataset to investigate the impact of each component in \framework. The results are shown in Table \ref{table:ablation_study}.

Beginning with Row 1, we observe that eliminating $s^\text{simul} (\cdot)$  Equation \eqref{eq:scoring_function} from the final calculation of relevance scores would lead to $10.3\%$ and $17.7\%$  MRR@10 and Recall@10 degradation in the MSMARCO Dev set, respectively. This is because the simultaneous relevance scoring function is based on set-based DocIDs which are constructed from a lexical approach. It can provide complementary relevance signals to the sequential DocIDs aiming at capturing semantic information. Row 2 also reflects the importance of combining lexical and semantic information for retrieval effectiveness. We find that solely using $s^\text{simul} (\cdot)$ for retrieval would result in a $27\%$ and $9.1\%$ decrease in terms of MRR@10 and Recall@10, respectively.

Based on Rows 3 and 4, we can infer that the more effective $M^{\text{seq}}$ can ultimately boost the effectiveness of the unified generative retrieval model. For instance, applying the seq2seq pre-training and multi-objective loss for fine-tuning can lead to a $1.0\%$ and $1.3\%$ enhancement on MRR@10 respectively. Seq2seq pre-training applies the DocID prediction task over the whole corpus, which can mitigate the issue of distribution shift between training and evaluation sets. The multi-objective loss is designed for sequentially decoding algorithms, such as beam search, used in generative retrieval models, which can reduce the risk of making the relevant prefixes discarded from the beam, especially in early decoding steps.

The results in Rows 5 and 6 imply that using $M^{\text{set}}$ or $M^{\text{seq}}$ alone would result in retrieval performance degradation. For example, only using $M^{\text{set}}$ and $M^{\text{seq}}$ reduce the MRR@10 by $19.6\%$ and $13.6\%$, respectively. Interestingly, when we linearly interpolate the lexical and semantic scores together for the final relevance score, we observe significant performance gain where the results are shown in Row 7. The simple post-hoc combination leads to $11.8\%$ and $6.2\%$ improvements over $M^{\text{set}}$ and $M^{\text{seq}}$ on the MSMARCO Dev set. This again emphasizes the effectiveness of combining lexical and semantic information for retrieval. That being said, the retrieval performance shown in Row 7 still lags behind the original model, which implies the effectiveness of integrating $M^{\text{set}}$ and $M^{\text{seq}}$ into a unified model with joint optimization. We observe that by joint modeling, the model can achieve $6.9\%$ and $13.0\%$ enhancement on MSMARCO@10 and Recall@10 respectively. 
%This is because the joint modeling can implicitly adjust the weights of set-based and sequential DocIDs for the final relevance scores, achieved via gradient descent on training data. Consequently, the end-to-end approach further boosts the retrieval effectiveness. 

Finally, as evidenced by Rows 8 and 9, \framework achieves superior results over $M^{sp}$ and $M^{ds}$, improving the MRR@10 by $1.9\%$ and $5.5\%$ respectively. Notably, this performance enhancement is accompanied by a significant reduction in memory usage - \framework requires $\times 10.8$ and $\times 7.7$ less memory compared to $M^{sp}$ and $M^{ds}$. The efficient memory utilization underscores the effectiveness of using set-based and sequential DocIDs in compressing the original embedding information near-losslessly and demonstrates the benefit of joint modeling them.

\begin{table}
    \centering 
    \caption{Ablation study results on MSMARCO Dev. Superscript $^\triangledown$ denotes significantly lower performance compared to \framework (t-test with Bonferroni correction, p\_value < 0.01). itp. stands for interpolation.}
    \scalebox{0.8}{\begin{tabular}
    {l!{{\color{lightgray}\vrule}}l!{{\color{lightgray}\vrule}}l!
    {{\color{lightgray}\vrule}}l!
    }
    \toprule
    & & & \multirow{2}{*}{\parbox{1.5cm}{ Index \\ Mem.(GB)}} \\
      & MRR@10 & Recall@10 &  \\
    \midrule
     \framework & .385 &  .670 & 3.27 \\
    \midrule
    1. w/o adding $s^{\text{simul}} (\cdot)$ & .349$^\triangledown$ & .614$^\triangledown$ & 0.50\\ 
    2. Only $s^{\text{simul}} (\cdot)$ for search & .303$^\triangledown$ 
    & .569$^\triangledown$ & 2.77 \\
    3. w/o seq2seq pre-training & .381$^\triangledown$ & .660$^\triangledown$ & 3.27 \\ 
    4. w/o multi-obj. learning & .380$^\triangledown$ & .663$^\triangledown$ & 3.27\\
    \midrule
    5. Only $M^{\text{set}}$ & .322$^\triangledown$ & .606$^\triangledown$ & 2.77\\ 
    6. Only $M^{\text{seq}}$ & .339$^\triangledown$ & .566$^\triangledown$ & 0.50 \\ 
    7. Linear itp. of $M^{\text{set}}$ and $M^{\text{seq}}$ & .360$^\triangledown$ & .593$^\triangledown$ & 3.27\\
    \midrule
    8. Only $M^{sp}$ & .378$^\triangledown$ & .667$^\triangledown$ &  35.28 \\
    9. Only $M^{ds}$ & .365$^\triangledown$ & .641$^\triangledown$ & 25.30 \\
    %10. Linear itp. of $M^{sp}$ and $M^{ds}$ & .389 & .679 & 60.58 \\
    \bottomrule
    \end{tabular}}
    \label{table:ablation_study}
\end{table}

\subsection{Analysis and Discussion}

\subsubsection{\textbf{The impact of beam size and number of selected sub-words}}
The selection of beam size and number of selected sub-words $m$ would affect the effectiveness and efficiency of the model. The large beam size can reduce the risk of pruning the relevant prefixes out of the beam and the large $m$ might improve the model expressiveness by extracting more lexical information from the document. However, it comes with the trade-off of increasing the query latency and index memory. To quantify that, we report the results of different settings of $m$ and beam size $k$ in Table \ref{tab:bow_and_beam_size}.
First, when $k$ is fixed, an increase of $m$ can show a trade-off in retrieval performance and resource utilization. For instance, at $k = 100$, the MRR@10 with $m=16$ is 0.355, while it increases to 0.390 when $m=128$, yielding $9.9\%$ enhancement. However, this gain comes at the cost of $\times3.58$ and $\times1.55$ increase in the index memory and query latency, respectively. 
Second, at a fixed value of $m$, employing a larger $k$ can enhance retrieval effectiveness. For instance, at $m=64$, increasing $k$ from 10 to 100 can improve MRR@10 by $1.6\%$ albeit at the expense of $\times5.68$ increase in query latency. It is noteworthy that performance degradation is less pronounced than that observed in RIPOR, as illustrated in Figure \ref{fig:RIPOR_perf_vs_bs}. This relative robustness can be attributed to the use of planning-ahead constrained beam search in \framework, which re-weights each prefix by a pre-stored document-level relevant score. Thus, it would facilitate more efficient retrieval without significantly compromising performance. Finally, we can observe that using simultaneous decoding is much more efficient than autoregressive generation for retrieval, hence does not introduce too much overhead over the original beam search algorithm.

\begin{table}[] 
    \centering
    \caption{The effectiveness and efficiency comparison with different $m$ and $k$ on MSMARCO Dev. The experiment is conducted on an 80GB A100 GPU. Simul. D. and Seq. D. stands for simultaneous and sequential decoding respectively. QL represents query latency (ms / query).}
    \scalebox{0.85}{\begin{tabular}{l!{{\color{lightgray}\vrule}}lllll}
        \toprule
         &  &  & \multirow{2}{*}{\parbox{1.5cm}{Index \\ Mem.(GB)}} & \multirow{2}{*}{\parbox{1.2cm}{Simul. D. \\ QL}} & \multirow{2}{*}{\parbox{1.2cm}{Seq. D. \\ QL}} \\ % & \multirow{2}{*}{\parbox{1.5cm}{Index \\ Mem.(GB)}} \\
        $m$, $k$  & MRR@10 & Recall@10 & & &  \\
        \midrule
        16, 10 & .342 & .577 & 1.30 & 20 &  44 \\
        32, 10 & .367 & .626 & 1.94 & 22 & 44 \\
        64, 10 & .379 & .641 & 3.27 & 25 & 44 \\
        128, 10 & .386 & .645 & 5.96 & 31 & 44 \\
        \midrule
        16, 100 & .355 & .620 & 1.30 & 20 & 250 \\
        32, 100 & .372 & .652 & 1.94 & 22 & 250 \\
        64, 100 & .385 & .670 & 3.27 & 25 &  250  \\
        128, 100 & .390 & .664  & 5.96 & 31 & 250 \\
        \bottomrule
    \end{tabular}}
    \label{tab:bow_and_beam_size}
\end{table}

\subsubsection{\textbf{Comparison between RIPOR and \framework}}
\begin{figure}
    \centering
    \includegraphics[width=.5\textwidth]{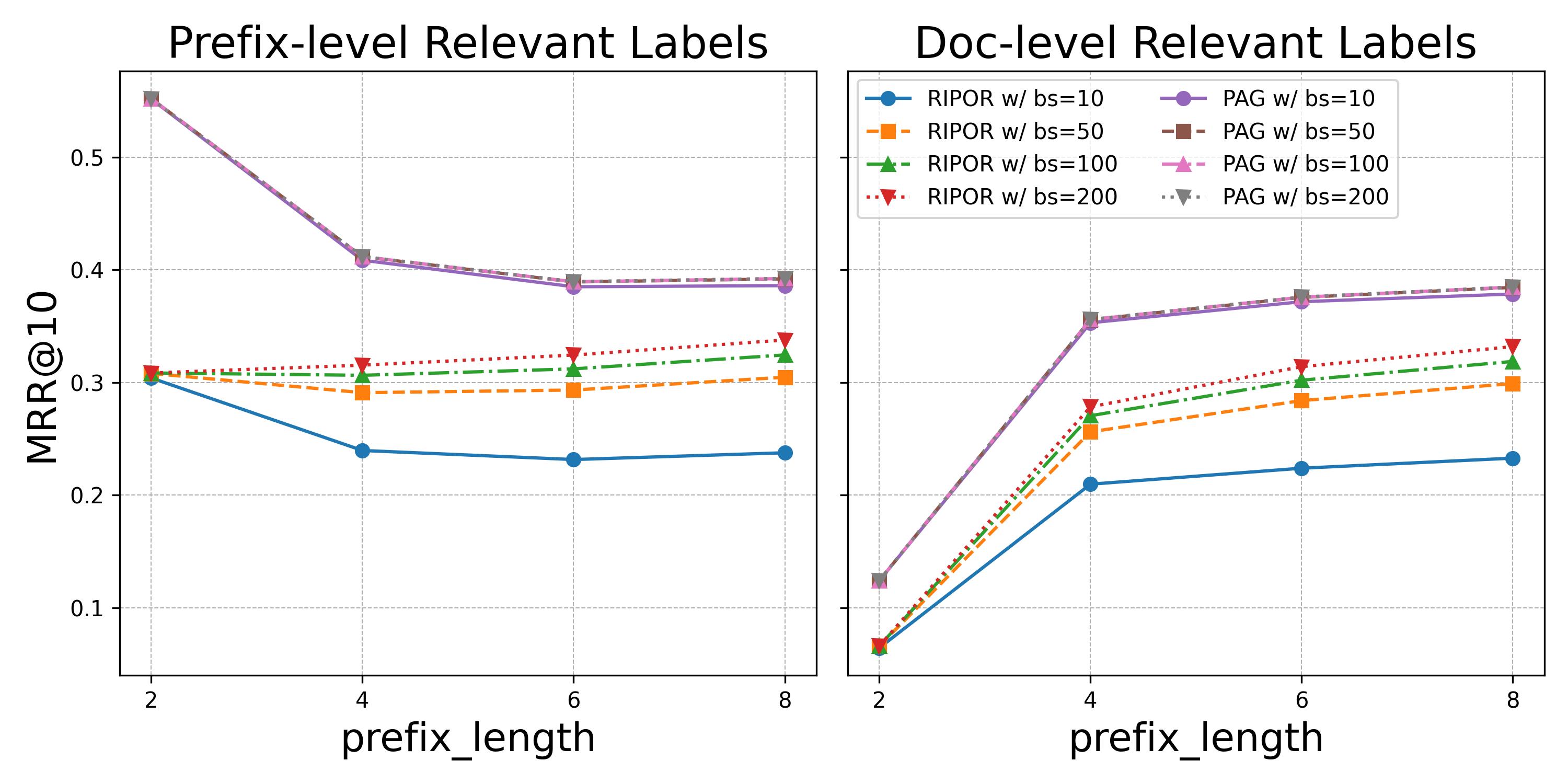}
    \caption{Results on MS MARCO Dev with different beam sizes on prefix-level and document-level labels.}
    \label{fig:perf_comp_diff_beam_size}
    \vspace{-.5 cm}
\end{figure}

We train a new RIPOR model with the same configurations of DocIDs used in \framework ($L=8$, $V=2048$). Other than the original document-level relevant labels, we also construct the prefix-level relevant labels where the prefix of a relevant document is also relevant to a given query. We compare RIPOR and \framework for different prefix lengths and beam sizes on the MSMARCO Dev set. The results are illustrated in Figure \ref{fig:perf_comp_diff_beam_size}. We find that \framework not only consistently outperforms in both prefix-level and document-level relevance labeling, but is also less sensitive to the beam size. MRR@10 for beam size greater than 10 have nearly identical performance across different prefix lengths. In contrast, we can always observe notable performance improvements with increased beam sizes in RIPOR. These findings indicate the effectiveness of incorporating document-level scores in constrained beam search. 

Additionally, our analysis reveals distinct patterns in prefix-level relevance labeling for RIPOR and \framework, as illustrated in the left sub-figure. In \framework, MRR@10 values have a monotonic decrease with longer prefix lengths. In contrast, RIPOR displays the opposite trends in MRR@10 except when the beam size is 10. It is because the use of simultaneous decoding for obtaining the document-level scores in \framework ensures high Recall@10 rates for early-stage relevant prefix retrieval. And making the relevant prefixes with higher ranks (better MRR@10) is easier in shorter lengths since shorter prefixes tend to be more dissimilar to each other, which reduces the challenges of distinguishing the relevant prefixes from hard negative prefixes \cite{ANN} in the retrieval system. 
%tend to be more coarse-grained and dissimilar to the irrelevant ones. 
This characteristic results in the decline of MRR@10 values as prefix lengths increase. Conversely, RIPOR lacks a similar early-stage Recall@10 performance. It relies on a larger beam size (>10) and longer, more expressive prefixes to achieve higher MRR@10 values.

\subsubsection{\textbf{The impact of combining set-based and sequential DocIDs}}
\begin{figure}
    \centering
    \includegraphics[width=0.5\textwidth]{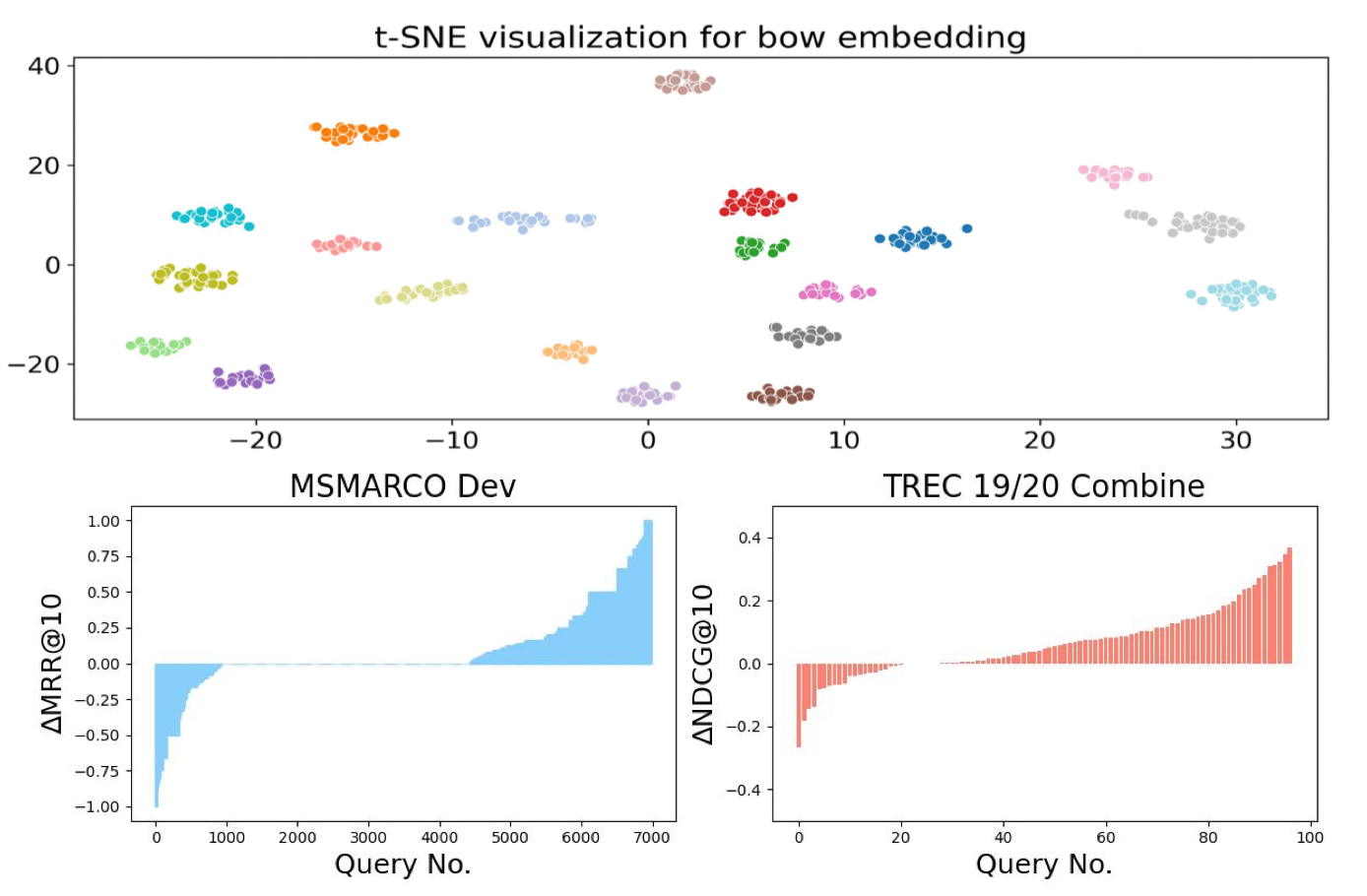}
    \caption{\textit{Above Figure}: clusters of relevant documents to 20 queries sampled from TREC-19/20, and the color indicates the query ID. \textit{Below Figures}: $\Delta$ MRR@10 on MSMARCO Dev and $\Delta$ NDCG@10 on TREC-19/20 between simultaneous+autoregressive decoding (\framework) and simultaneous decoding alone for each query.
    }
    \label{fig:perf_diff_by_query}
    \vspace{-.5 cm}
\end{figure}

We can alternatively view \framework as a retrieve-then-rerank model, in which it first retrieves the top promising documents using set-based DocIDs and then gradually refines the retrieved document scores based on the newly generated next token by sequential DocIDs.  
In Table \ref{table:ablation_study}, we already show that only using set-based DocIDs for search can achieve $.303$ in terms of MRR@10, while in this section, we qualitatively demonstrate the retrieval effectiveness by investigating the quality of set-based DocIDs. For this, we randomly sampled 20 queries from the combining sets of TREC 19/20, and selected all the relevant documents to these sampled queries. For each document $d$ with the set-based DocID $t^d = \{t^d_1, \ldots t^d_m \}$, we obtain corresponding T5-embedding (learned in \framework) for each token and denoted them as $\{\mathbf{e}^d_1, \ldots \mathbf{e}^d_m\}$, then each document embedding can be computed as $\mathbf{e}_d = \frac{1}{m} \displaystyle\sum_{i=1}^m \mathbf{e}^d_i$. We apply the T-SNE \cite{t-sne} to the document embeddings for dimension reduction and visualize them in the above figure of Figure \ref{fig:perf_diff_by_query}. According to the figure, we observe that nearly all documents with the same label (relevant to the same query) can be clustered together. This demonstrates that the set-based DocID can extract useful tokens from each document's content for capturing relevance-based information and enhancing retrieval performance.

To better understand the re-ranking effect of combining set-based and sequential DocIDs for computing relevance scores. We compare the performance difference between the original \framework and the variant that only uses set-based DocIDs for retrieval. We report the $\Delta$MRR@10 in the MSMARCO and the $\Delta$NDCG@10 in the TREC-DL 19\&20 sets respectively for each query, and the results are illustrated in the below sub-figures of Figure \ref{fig:perf_diff_by_query}. For the sake of space, we merged the query sets of TREC-DL 19\&20 in this experiment. We observe from the plots that most queries can either benefit or at least not be harmed by joint scoring.
We acknowledge that not all queries benefit from the joint scoring. Specifically, approximately 1000 out of 6980 queries in MSMARCO Dev and 20 out of 97 queries in TREC-DL 19\&20 notice a decline in performance. This could be attributed to combined scoring potentially introducing biases that negatively affect queries that are better suited to lexical matching, typically captured by set-based DocIDs.

%\subsubsection{\textbf{Efficiency and effectiveness trade-off for different generative retrieval models}}
%The choice of beam size would affect the effectiveness and efficiency of the generative retrieval models. We explore beam sizes ranging from $\{ 10, 50, 100, 200, 500, 1000 \}$ and study their impact on MRR@10 and query latency for various generative retrieval models in Figure \ref{fig:diff_gr_perf_vs_latency}. 
%Initially, we observe divergent trends in MRR@10 relative to query latency among different generative retrieval models. For instance, LTRGR, which employs the FM-index \cite{SEAL} for searching, experiences a performance drop as query latency increases. In contrast, models using constrained beam search, like DSI-QG, RIPOR, and \framework, either show improved performance or maintain consistent results with larger query latency.
%Moreover, \framework stands out in both retrieval effectiveness and efficiency. For example, it outperforms RIPOR, achieving better results with query latency that is $100$ times faster. This highlights the significant impact of integrating the document-level relevant scoring function and the joint modeling. 

%\begin{figure}
%    \centering
%    \includegraphics[width=0.45\textwidth]{imgs/diff_gr_comp_perf_vs_lantency.jpg}
%    \caption{The MRR@10 w.r.t query latency (dependent on the selection of beam size) for different GR models in the MSMARCO Dev set. Each point represents a certain beam size ranging from $\{10,50,100,200,500,1000\}$.}
%    \label{fig:diff_gr_perf_vs_latency}
%\end{figure}

\section{Conclusions and Future Work}
In this paper, we proposed a novel optimization and decoding framework for generative retrieval. We introduced simultaneous decoding for efficient document-level score estimation and used it to guide autoregressive decoding. We create the set-based DocIDs under the bag-of-words assumption and sequential DocIDs based on the relevance-based document representations to support simultaneous and autoregressive decodings, respectively. We additionally introduced a three-stage training pipeline for gradual adaptation of the model to joint decoding. Our experiments demonstrated substantial improvements compared to state-of-the-art generative retrieval baselines, in terms of both efficiency and effectiveness. Looking ahead, we aim to further optimize the model's efficiency and scale it up to billion-scale datasets. We also look forward to integrating the framework into other knowledge-intensive tasks, such as open-domain question answering.

\section{Acknowledgment}
This work was supported in part by the Center for Intelligent Information Retrieval, in part by Lowe’s, and in part by an Amazon Research Award, Fall 2022 CFP. Any opinions, findings and conclusions or recommendations expressed in this material are those of the authors and do not necessarily reflect those of the sponsor.
% \hamed{write here. Especially reiterate your key findings.}

\bibliographystyle{ACM-Reference-Format}
%\balance
\bibliography{sample-base}
\end{document}